# A Survey on Federated Learning and its Applications for Accelerating Industrial Internet of Things

Jiehan Zhou
jiehan.zhou@ieee.org
University of Oulu

Shouhua Zhang
Shouhua.zhang@oulu.fi
University of Oulu

Qinghua Lu
qinghua.lu@data61.csiro.au
CSIRO, Data61 13 Garden Street Eveleigh SYDNEY

Wenbin Dai
w.dai@ieee.org
Shanghai Jiao Tong University

Min Chen
minchen@ieee.org
Huazhong Univ. of Science and Technology, China

Xin Liu
lx@upc.edu.cn
China University of Petroleum Huadong

Susanna Pirttikangas
susanna.pirttikangas@oulu.fi
University of Oulu

Yang Shi
yshi@uvic.ca
University of Victoria

Weishan Zhang
zhangws@upc.edu.cn
China University of Petroleum Huadong - Qingdao Campus

Enrique Herrera-Viedma
viedma@decsai.ugr.es

*Abstract*—Federated learning (FL) brings collaborative intelligence into industries without centralized training data to accelerate the process of Industry 4.0 on the edge computing level. FL solves the dilemma in which enterprises wish to make the use of data intelligence with security concerns. To accelerate industrial Internet of things with the further leverage of FL, existing achievements on FL are developed from three aspects: 1) define terminologies and elaborate a general framework of FL for accommodating various scenarios; 2) discuss the state-of-the-art of FL on fundamental researches including data partitioning, privacy preservation, model optimization, local model transportation, personalization, motivation mechanism, platform & tools, and benchmark; 3) discuss the impacts of FL from the economic perspective. To attract more attention from industrial academia and practice, a FL-transformed manufacturing paradigm is presented, and future research directions of FL are given and possible immediate applications in Industry 4.0 domain are also proposed.

*Index Terms*—Federated learning, Internet of Things, Industry 4.0, deep learning, edge computing

## I. Introduction

Google first proposed [1] FL to aggregate distributed intelligence without compromising data privacy security. The increasing attention of FL comes from the combined force of emerging new technologies with applications. Although Industry 4.0 was proposed in 2013 [2] and Internet of Things (IoT) is being widely applied in mobile services. There are few reports on applying large-scale data and deep learning (DL) to implement large-scale enterprise intelligence. One of the reasons is lack of machine learning (ML) approaches which can make distributed learning available while not infringing the user's data privacy. Clearly, FL trains a model by enabling the individual devices to act as local learners and send local model parameters to a federal server (defined in section 2) instead of training data. This gives a clear advantage in terms of privacy-oriented industrial applications. Another key advantage is that FL does not need large data-sets to be moved to a central repository (edge/cloud), it avoids known problems related to the sink node congestion/overloading. Another advantage of FL is to give small and medium-sized enterprise (SMEs) an opportunity to make full use of intelligence, which might be lack of large sets of data and more eager to apply FL into balancing data intelligence and proprietary for promoting innovation and enhancing competitiveness.

There have been several surveys on FL. For example, Yang *et al.* [3] made a seminal survey that introduces the basic concepts in FL and a secure FL framework. Aledhari *et al.* [4] provided a study of FL with an emphasis on enabling software and hardware platforms, protocols, real-life applications and use-cases. Li *et al.* [5] discussed the unique characteristics and challenges of FL, provided a broad overview of current approaches, and outlined several directions for future work. Lo *et al*. [6] performed a systematic literature review on FL from the software engineering perspective. Li *et al.* [7] conducted a review of FL systems, introduced the definition of FL systems and analyzed the system components. Mothukuri *et al.* [8] provided a study concerning FL's security and privacy aspects and outlined the areas which require in-depth research and investigation. The early reviews introduced the basic concepts and optimization models of FL. Recently, related platforms and tools are developed, incentive mechanisms are considered, and benchmarks and personalized FL are added as well. The FL architecture needs to be updated as well to accommodate the increasing FL research and development. Meanwhile, it is noted that most FL pioneers come from the fields of the computer and information communication community, and may not put enough emphasis on the communication with industrial engineering, which seriously hinders the application of FL on industrial Internet of Things (IIoT) and the development of IIoT.

Therefore, we revisit this hot topic from the perspective of promoting Industry 4.0, incorporating the consideration from the practice of industrial big data [9] and edge computing [10]. Our contribution in this survey lies in two aspects: a comprehensive investigation of the state of the art on FL, including fundamental and applied research; attracting and aggregating attentions from informatics and industrial expertise to advance the application of FL into Industry 4.0 by presenting our insights on promoting industrial data protection and intelligence.

The remainder of the paper is organized as follows. Section II goes over the origin and development of FL, defines the terminology used in FL and this paper, and describes the FL mechanism in our terminology. Section III reviews the state of the art on fundamental FL and future opportunities. Section IV presents the FL-transformed manufacturing paradigm and reviews the state of the practice on FL and future opportunities, specially in Industry 4.0. Section V concludes the paper and presents the insights for advancing FL studies.

## II. FEDERATED LEARNING

### A. Evolution of FL

FL is one of the future generation of artificial intelligence (AI), and it is also based on the latest stage of information communication technology (ICT) and new hardware technologies. After AlphaGo successfully defeated professional Go players in 2015, AI once again attracted worldwide attention [11]. ML is a part of AI. ML algorithms build models based on sample data (called "training data") in order to make predictions or decisions without explicit programming [12]. ML and Data Mining (DM) have a lot of overlap, but ML focuses on prediction based on learned information from training data, while data mining focuses on discovering unknown information in the data. DL is a part of ML based on artificial neural networks with representation learning. Learning can be supervised, semi-supervised or unsupervised. DL has various learning structures, such as deep neural networks, recurrent neural networks and convolutional neural networks. They have been used in machine vision, speech recognition, natural language processing, etc., where their produced results can be comparable to and surpass the performance of human experts in some cases [13][14]. Distributed machine learning (DML) is a multi-node-based ML where the master node cooperates with each slave node to train a model in parallel to improve learning performance from large amounts of data [15][16]. This traditional "centralized" distributed learning still has some drawbacks [17]: low efficiency with high transmission cost and lack of privacy preservation which significantly reduce application levels of DL in domains, for example, manufacturing. Besides, limitations on providing enough training data and computing power prevent many industries from adopting ML. Also, most industrial manufacturers would not share their data for security and privacy reasons. FL is a part of DML, which is defined in the next section.

### B. Definitions and Terminologies

FL, also called federated machine learning, is an ML framework that can effectively make use of data and perform ML without having to share local data. Based on the mathematical formulation given in [3][7], we refine the following conditions 1) and 2) for describing the accuracy of an FL for facilitating the following discussions. The terms relevant to FL are listed in Table 1. Assume that there are $N$ different learners $L$ who aim to train the FM together. Each learner is denoted by $L_i$, where $i \in [1,N]$. $D_i$ denotes the raw data owned by $L_i$ and participated in FL. For a non-federated setting, put all the data together and use $D = D_1 \cup \ldots \cup D_N$ to train a model $M_{\text{center}}$. The predictive accuracy of $M_{\text{center}}$ is denoted as $A_{\text{center}}$. For another non-federated setting, each learner $L_i$ trains a local model $LM_i$ with $D_i$ separately. The predictive accuracy of $LM_i$ is denoted as $A_i$. For the federated setting, all the learners collaboratively train a model $M_{\text{fed}}$ while each learner $L_i$ protects its own data $D_i$ based on its privacy constraint. The predictive accuracy of $M_{\text{fed}}$ denoted as $A_{\text{fed}}$ should be very close to $A_{\text{center}}$. Formally, let ε be non-negative real number; if

$$| A_{\text{fed}} - A_{\text{center}} | < \varepsilon \qquad (1)$$
$$A_{\text{fed}} > A_i \ ( i \in [1,N]) \qquad (2)$$

then we say that the algorithm for FL has ε accuracy loss. Let $SI_i$ denote the sample id space of $D_i$. Let $X_i$ denote the feature space of $D_i$. Let $Y_i$ denote the label space of $D_i$. So, we use ($SI_i$, $X_i$, $Y_i$) to represent $D_i$.

FL itself does not guarantee data privacy. After each round of training, learner $L_i$ will share the local model $LM_i$, and other learners or organizer can reconstruct part of $L_i$'s information based on $LM_i$. We propose a privacy measurement method based on reverse reconstruction. Suppose $X_i = [ x_i(1), x_i(2),\ldots, x_i(M_i)]$ and $M_i$ is the feature number of $X_i$. The learner $L_i$ reversely reconstructs $X_i$, which is expressed in Equation (3). The learner $L_j$ or organizer reversely reconstructs $X_i$, which is expressed in Equation (4). When the $L_j$ or organizer has no data, it can randomly initialize a dummy $X_j$ and $Y_j$.

The privacy measurement of $L_i$ reconstructing its own original data with $LM_i$ is expressed in Equation (5), and the privacy measurement of others reconstructing the original data of $L_i$ with $LM_i$ is expressed in Equation (6). Equation (3) is the benchmark that has the highest similarity.

$$\hat{X}_{(i \to i)} = \{x_{(i \to i)}(k) | Y_i \xrightarrow{LM_i(k)} x_{(i \to i)}(k), k = 1, 2, \cdots, M_i \quad (3)$$

$$\hat{X}_{(j \to i)} = \{x_{(j \to i)}(k) | Y_j \xrightarrow{LM_i(k)} x_{(j \to i)}(k), k = 1, 2, \cdots, M_j \quad (4)$$

$$Privacy(X_{(i)}, \hat{X}_{(i \to i)}) \coloneqq 1 - d(X_{(i)}, \hat{X}_{(i \to i)}) \quad (5)$$

$$Privacy(X_{(i)}, \hat{X}_{(j \to i)}) \coloneqq 1 - d(X_{(i)}, \hat{X}_{(j \to i)}) \quad (6)$$

where $d$ denotes the measurement method, such as Euclidean distance. The discussion on the calculation method of FL privacy and $d$ is beyond the scope of this paper.

### C. FL Mechanism

We first take the industrial equipment health monitoring as a typical example of HFL to describe the procedure of FL (Figure 1) with the above terminologies. It is a common centralized architecture. The decentralized architecture is described in Section III. Suppose that $N$ companies are participating in FL. That is, there are $N$ learners. The basic learning steps are as follows [19]:

1) The organizer chooses an FM and initializes its parameters.
2) The organizer calls FM transmitter to send FM to all the learners participating the learning.
3) FM receiver of $L_i$ ($i \in [1,N]$) receives and stores it.
4) $L_i$ calls the trainer to train $LM_i$ with local data and FM.
5) $L_i$ calls LM transmitter to send the $LM_i$ to the organizer.
6) The LM receiver receives each $LM_i$.
7) The optimizer updates FM with the aggregation algorithm and the received LMs.
8) Repeat the above step 2 to step 7 until convergence.

Second, we describe a typical example of applying VFL as follows. Suppose that dealer A and company B want to build a sales forecast model for company B's products based on the data owned by both parties. We denote dealer A as $L_A$ and company B as $L_B$. We denote the sales data owned by dealer A as $D_A$. $D_A$ can be represented by ($SI_A$, $X_A$, $Y_A$). We denote the data on product processing owned by company B as $D_B$. $D_B$ can be represented by ($SI_B$, $X_B$). The basic learning steps are as follows [19]:

1) $L_A$ and $L_B$ align the sample data with samples' id.
2) $L_A$ and $L_B$ choose an FM. $L_A$ initializes part of the parameters of FM according to $D_A$, that is $LM_A$. $L_B$ initializes part of the parameters of FM according to $D_B$, that is $LM_B$. $LM_A$ and $LM_B$ make up all the parameters of FM.
3) $L_B$ calls LM trainer for a round of training and sends the result $M_B$ to $L_A$ and $L_A$ calls LM trainer for a round of training and sends the result $M_A$ to $L_B$ [3].
4) $L_A$ calculates the loss $LS$ and the gradient $G_A$ with $M_A$ and $M_B$. $L_B$ calculates the gradient $G_B$ with $M_A$ and $M_B$.
5) $L_A$ updates $LM_A$ with $G_A$ and $L_B$ updates $LM_B$ with $G_B$.
6) Repeat steps 3-5 until convergence.

Third, a typical example of applying FTL is as follows. Transfer learning aims at shifting knowledge from existing domains to a new domain. When dealer A has sold a small number of company B's products (Figure 1), dealer A and company B still want to build a sales forecast model for company B based on the data owned by both parties. The learning process of FTL is similar to VFL, except that the details of the intermediate results exchanged between A and B are changed [3][20].

These three kinds of FL mechanisms can help all participants in the above example make full use of the original data of federation members to realize intelligent sharing based on large-scale data, while protecting the privacy of the original data.

### III. THE STATE OF THE ART ON FL

In this section, we present a comprehensive review and analysis of the fundamental studies on FL in the past two years，excluding the studies of integrating learning paradigms such as unsupervised learning [21][22].

Table I
The terminologies in FL

| Term | Denotation |
|---|---|
| data owner | The user who owns data collected from *clients*. |
| learner | The *data owner* that contributes data and particulates in *LM* training. |
| organizer | The user is trusted by *learners*, who coordinates FL system, coordinating learners to train FM via servers and managing *FMs*. It is also called *coordinator/collaborator* or *third party* in literature. |
| beneficiary | The user who uses *FM*, may not be a *learner*. |
| local model (LM) | The model is iteratively trained by learners using local data based on *FM model* offered by *organizer*. |
| federal model (FM) | The model that is iteratively optimized by *optimizer* based on *LMs* contributed by *learners*, which is also called *global model* in literature [18]. |
| optimizer | A general term for the algorithm which is used to collectively synthesize and optimize *LMs*. |
| client | A tool platform on *learner* side, including hardware and software, mainly collecting raw data, training *LMs*, transmitting *LMs*, receiving *FMs*, storing *LMs* and *FMs*. |
| LM trainer | A general term for the algorithm for training the model using local data. |
| server | The platform for operating system executing FL, including hardware and software, and mainly running *optimizer*, receiving *LMs*, transmitting *FMs*, storing *LMs* and *FMs*. |
| LM transmitter | The technique transmits *LMs* between *client* and *server* following Service Level Agreements (SLAs). |
| LM receiver | The module receives *LMs* from *learners*. |
| data collector | The module collects raw data, e.g., via the internet of things. |
| FM transmitter | The technique transmits *FMs* between *server* and *client* following Service Level Agreements (SLAs). |
| FM receiver | The module receives *FMs* from the *server*. |
| LM and FM storage | The database stores, replicates, and updates *LMs* and *FMs*. |
| centralized FL (CFL) | FL is centralized and controlled for building *FM* by an *organizer* with all *LMs* from *learners*. |
| decentralized FL (DFL) | Some *learners* are responsible for collecting *LMs* and building *FMs* in FL. |
| aggregation algorithm (AA) | The common algorithm used in FL. |
| horizontal federated learning(HFL) | HFL: $X_i = X_j$, $Y_i = Y_j$, $SI_i \neq SI_j$ ($i \neq j$) |
| vertical federated learning(VFL) | VFL: $X_i \neq X_j$, $Y_i \neq Y_j$, $SI_i = SI_j$ ($i \neq j$) |
| federated transfer learning(FTL) | FTL: $X_i \neq X_j$, $Y_i \neq Y_j$, $SI_i \neq SI_j$ ($i \neq j$) |

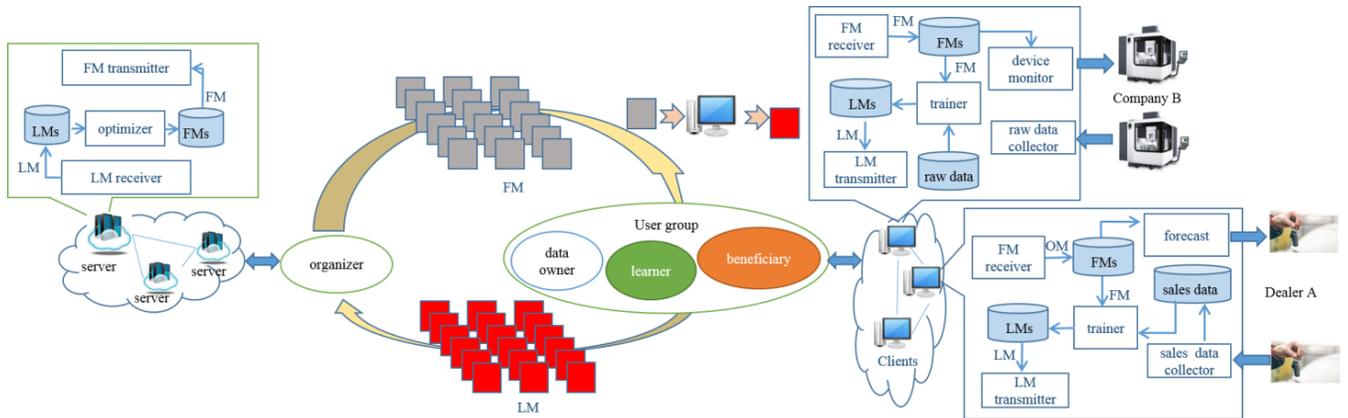
Figure 1. The general FL implementation platform

## A. Fundamental research

*Data Partitioning (DP)*

Data partitioning is significant in the learning process. HFL is the most commonly adopted approach in both cross-device and cross-silo scenarios where data can hardly be centralized due to privacy or legal concerns. Cross-device FL aims to train application-centered models from the collaboration of a large-scale distributed network, with a massive number of smart devices, whilst cross-silo FL does not allow to share data between involving organizations [23][24]. In the cross-device setting, HFL handles the situation in product/service design when data analysis is integrated as a feature of the personalized product but with data privacy concerns, e.g., Google's mobile virtual keyboard prediction [25], device failure detection [26]. In the cross-silo setting, HFL has been applied to the case when organizations share the same ML problems but under restricted data sharing policies, e.g., COVID-19 detection using diagnostic images from different medical institutions [27]. VFL is usually considered in the cross-silo setting when two organizations have the shared set of sample data but different ML objectives [3][28][29][30], e.g., between a bank and an insurance company located in the same city, or between smart refrigerators and smart air conditioners produced by different manufacturers. In the cross-silo setting, when the participating organizations (usually only two organization involved) only have the partial shared set of sample space or feature space, transfer learning techniques [31] can be adopted in FL to train models collaboratively [31][32][33][34].

*Privacy Preservation (PP)*

Data privacy is still the major challenge of FL since it is possible to leak private information through analysis on updates of local model parameters or gradients [6][35]. There are mainly two ways to address this issue: secure multiparty computation and differential privacy. Homomorphic encryption is a technique to realize secure multiparty computation, which only allows the central server to conduct homomorphic computing based on the encrypted local model updates [36][37]. Trusted Execution Environment can empower the detection of dishonest actions (e.g., tampering with client models, delaying local training, etc.), to guarantee the integrity of FL processes [38]. Differential privacy is often used to protect client data privacy by adding noise to model parameter data sent by each client [39]. Additionally, a hybrid approach combining secure multiparty computation with differential privacy is explored in [40]. Recently, Blockchain is used to share data generated and used in the model training, and clients can control the access to shared data [30]. Specifically, a directed acyclic graph is incorporated to improve the efficiency of data sharing, while an asynchronous FL scheme can minimize the total cost [41]. Further, model updates can be directly exchanged and verified on-chain [42], which needs to separate clients into different groups and each group is assigned a miner to gather the model updates. Another option is to store original global updates off-chain, and only save the pointer of the global updates to improve efficiency [43].

*Model Optimization (MO)*

Federated averaging (FedAvg) algorithm is the first and most well-known algorithm proposed by Google [44], which aggregates local model updates sent from clients for a federal model. However, the FedAvg algorithm fails to achieve a satisfactory model and system performance when the datasets produced by different clients are not independent and identically distributed (Non-IID) and the communication cost is high [45]. To solve this issue, particularly in the context of industrial IoT [46], the algorithm optimization plays an important role in FL. Centroid distance based FedAvg approach is proposed to consider the centroid distance between each class as a metric of data heterogeneity and take it into the updated averaging [26]. Bounds expanding is used to handle data skew, which extends the bounds of each dataset by exchanging some data to make the data distribution similar [47]. A self-organized FL framework is proposed in [48], where the server has the capability of recognizing heterogeneity and scheduling a stable collaboration plan for client selection. An optimal tuning on the distributed training set is achieved by a collaborative teaching approach to train models on the optimal tuning for better performance [49].

FL itself adopts a distributed topology via collaboration among participating clients in ML. However, it still maintains a settled centralized architecture where a server is required for model aggregation and distribution. Some studies investigate improving further decentralization to get over the restrictions of a fixed server-client architecture. [50] removes the central server, and clients need to communicate with each other for model update in each round. The whole network can be split into several subsets and each one is responsible for a certain part of the expected model [51]. Gossip learning is also considered as a decentralized alternative of FL [52]. In addition, blockchain can be exploited as a component to enable decentralized infrastructure in FL [53][54].

*Local model transportation (LMT)*

As model updates are uploaded for aggregation by client devices that have slow connections to the server, it is valuable to improve the communication efficiency between clients and the server. The initial research focused on the synchronous update scheme [44]. In each epoch, some clients are randomly selected, and the server sends the current federal model to each of these clients. Then, each client performs local training based on the federal model and its local dataset, and sends updates to the server. The server then updates the federal model with these updates, and the process repeats. Asynchronous aggregation is used to update the federal model asynchronously to reduce the response time from the server [27][45][55][56]. Sparse ternary compression is proposed to satisfy high-frequency and low-bit width communication, which compresses both upstream and downstream communications, and enables optimal Golomb encoding of the weight updates [57]. The Lyapunov optimization-based load balancing is used to reduce communication overhead [58]. To decrease the times of sending updates that are irrelevant to the improvement of the federal model, each client receives a global tendency of model updating as feedback and checks its updates with the global tendency. If client model updates do not align with the global tendency, the client will not upload the upgrades to the server [59].

*Personalization (Per)*

The concept of personalized FL emerged to reduce heterogeneity and preserve the high-quality of client contributions. In order to tackle the challenges of device heterogeneity, statistical heterogeneity and model heterogeneity, an effective method is to implement personalization in device, data and model levels to reduce heterogeneity and obtain high-quality personalized models for each device. Researchers from Google proposed three approaches to FL personalization [60]: 1) user clustering where the clients are divided into different groups and collectively train a model for each group; 2) data interpolation in which some data is shared as global data, and a model is trained using both local and global data; 3) model interpolation that combines the learned and optimized models. Based on these methods, a synergistic cloud-edge framework is proposed, which allows each client to offload its computationally intensive learning task to the edge [61]. Besides the mixture of local and federal models, the efficient optimization of communication shows better performance on convergence [62]. Furthermore, the Model Agnostic Meta Learning framework is similar to the personalization of FL, and can be used for the interpretation of existing FL algorithms [63][64].

*Motivation Mechanisms (MM)*

Incentive mechanisms are considered as an effective way to ensure the long-term stability of FL and motivate clients to provide learned models with higher quality. Data size and quality can be considered in the design of incentive mechanisms [26]. With a limited budget, incentives given to clients can be designed by computing solutions for payoff-sharing with instalment [65]. Furthermore, the theory of Stackelberg game can be applied, in which the central server is a buyer for training service provided by clients [66]. Clients can decide the CPU power for gradient calculation based on the given incentive. To ensure both clients' enthusiasm and the quality of the aggregated model with diverse metrics, three kinds of fairness (i.e., contribution fairness, regret distribution fairness, and expectation fairness) are taken into account, to optimize the collective utility while minimizing corresponding inequalities [65]. A reputation mechanism is proved as a feasible way to ensure the trustworthiness of clients, which can record reputation histories on blockchain for tamper-resistance properties in a decentralized manner [67]. Blockchain can be also leveraged for the voting of clients' rewards that clients chosen in the current round need to vote for the previous model updates [68].

*Platforms and Tools (PT)*

As FL involves multiparty computation to gather model updates for optimization, developing a user-friendly platform can ease the operations and maintenance [69][70]. There are several mature FL platforms from the industry, including Federated AI Technology Enabler (FATE), TensorFlow Federated (TFF), OpenMined PySyft, PaddleFL, LEAF. Further, Flower is an open-sourced framework for practitioners to conduct experiments and implement their federated learning schemes [71].

*Benchmark (Ben)*

In edge computing scenarios, various devices and cloud servers are coordinated to maintain communication and data analysis, which requires computing power, data storage and bandwidth. Therefore, a unified testbed is required to support the development of FL systems in complicated scenarios. Edge AIBench is proposed by BenchCouncil for edge AI benchmarks [72].

B. *Future opportunities*

The emergence of FL has brought many opportunities to ML for IIoT, but it also faces more challenges. According to the application of fundamental research in FL for IIoT, we emphasize some future works that deserve further investigation in the following.
- **Privacy preservation**. Quantifying data privacy exposure

has not been fully studied in existing studies. The current research focuses on learning accuracy and does not study data privacy measurement. We believe that it is necessary to establish a mechanism to evaluate data privacy exposure like model accuracy in the future. Meanwhile, learners actually have different needs for data privacy, but it is currently limited to privacy protection at the same level.

- **Model evaluation criteria**. The current model evaluations are all based on a third party and lack a universal and unified evaluation standard, such as representative data sets for evaluation, load, etc. Therefore, the establishment of a benchmark for FL is an important direction.
- **Personalization**. The storage, computing, and communication capabilities of each client device in the federal network may vary due to differences in hardware, network connections, and power. Due to connectivity or energy constraints, it is also common for client devices to lose communication during iteration. These bring challenges to straggler mitigation and fault tolerance. The differences in equipment and data collection methods violate the independent and identically distributed assumptions, and may increase the complexity of problem modeling and theoretical analysis.
- **Incentive mechanism**. There is currently a lack of effective incentive mechanisms in FL, such as contracts for more work, more rewards.
- **Data distribution**. At present, most researches focus on HFL, but there are rather few well developed algorithms for VFL. However, VFL applications are common in industry involving multiple organizations.
- **Local model transportation**. There are already methods that can significantly reduce communication costs with little impact on training accuracy. It is unclear whether communication costs can be further reduced, and whether these methods can provide the best trade-off between communication and accuracy in FL.
- **Model optimization**. When the client updates the federal model in an asynchronous or lock-free manner, error convergence analysis is an open and challenging problem. Asynchronous methods can be difficult to combine with technologies such as differential privacy or secure aggregation. Standard FL is usually hosted and operated by a central server, which is somehow criticized for such a centralized mode. Higher level of decentralization can be further studied to alleviate this plight, for the fairness in possible coordination among multiple parties within Industrial IoT.
- **Platform and tools**. A comprehensive platform is needed for covering the functional requirements from raw data processing, model storage, model training, model transportation, aggregation algorithms, data privacy preservation, incentive mechanism, personalization, etc.
- **Security**. FL is still vulnerable to some attack models such as inference attack and poisoning attack [17]. Adversaries upload malicious updates to the server for aggregation, which may have a significant impact on the federal model. Curious or malicious servers can easily use the shared computing power to build malicious tasks in the federal ML model. Adversaries can partially reveal the training data of each participants' original training data according to the local models uploaded by them. Emerging challenges still exist when applying FL to IIoT.

## IV. FL-BASED APPLICATIONS

Figure 2 illustrates how FL could be applied to product life cycle management under the concept of Industry 4.0. FL expects to be widely applied in harvesting powerful intelligence in enhancing product life cycle management (PLCM) with the deep implementation of Industry 4.0. In the product R&D phase, market demand discovery and product innovation can be devised based on FL. In the production phase, FL paves the way for making use of industrial big data across enterprises to leverage effective and efficient utilization of manufacturing resources of energy, device, manpower, tool, etc. In the marketing phase, FL can improve product marketing efficiency with the analysis of market data contributed by federal members. The FL-transformed manufacturing paradigm shows a quite broad spectrum to utilize FL.

### A. The state of practice

According to the conditions for applying FL described in Section 2, we provide our analysis and summary of important FL applications, according to different application areas that were reported in the past two years. There are few applications spreading over PLCM. More attention should be paid on utilizing FL in IIoT. Table II summarizes the related literature.

### FL for IIoT

Zhang *et al*. [26] proposed an FL method based on blockchain to detect device failures in IIoT. A platform architecture of FL system based on blockchain is designed, which supports verifiable integrity of client data. Each client periodically creates a Merkle tree where each leaf node represents a client data record and the root is stored on the blockchain. Moreover, a new centroid distance weighted

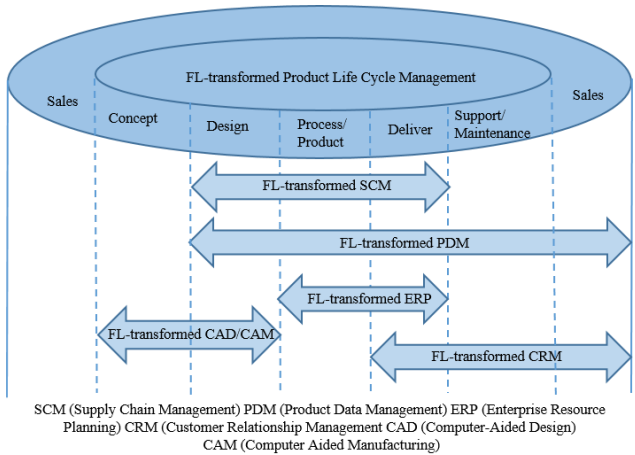

Figure 2. FL-transformed manufacturing for PLCM

Table II.
The distribution of fundamental research in the investigated applications

| Application | | MO | LMT | DP | PP | MM | Per | PT | TM | Ben |
|---|---|---|---|---|---|---|---|---|---|---|
| AD | FTRL[102] | CFL, FedAvg | Async,Sync | FTL | Bas | N/A | N/A | N/A | DDPG | SM |
| IIoT | [79] | CFL, EFA | Sync | HFL | Bas | N/A | N/A | N/A | LSTM | SM |
| | [26] | CFL,CDW_FedAvg | Sync | HFL | Enc | Yes | N/A | Leaf | DNN | SM |
| | [73] | CFL,FedAvg | Sync | HFL, VFL | Bas | N/A | N/A | N/A | SVM,RF | SM |
| | [74] | CFL,FedAvg | Sync | HFL | Bas | N/A | N/A | N/A | CNN | SM |
| | AMCNN-LSTM[75] | CFL,FedAvg | Sync | HFL | Bas | N/A | N/A | PySyft | CNN LSTM | SM |
| | [76] | CFL, EFA | Async | HFL | Bas | N/A | N/A | N/A | DQN | SM |
| | DeepFed[77] | CFL, ParaAggregate | Sync | HFL | Enc | N/A | N/A | Flask | CNN,GNU | SM |
| | LFRL[33] | CFL,KFA | Async | FTL | Bas | N/A | N/A | N/A | CNN | SM |
| | FIL[78] | CFL, KFA | Async,Sync | FTL | Bas | N/A | MH | N/A | CNN | SM |
| SB | DÏOT[100] | CFL, EFA | Sync,CEC | HFL | Bas | N/A | MH | TF | GRU | SM |
| | [101] | CFL, FedAvg | Sync, CEC | HFL | Bas | N/A | N/A | N/A | CNN | SM |
| SE | FEDL[99] | CFL, FedAvg | Sync | HFL | Bas | N/A | N/A | TF | DNN | SM |
| SC | FedSem[98] | CFL, FedAvg | Sync | HFL | Bas | N/A | StH | TF | CNN | SM |
| RS | FCF[87] | CFL, FCF | Sync | HFL | Bas | N/A | N/A | N/A | CF | SM |
| | [88] | CFL, FedAvg | Async | HFL | Bas | N/A | N/A | N/A | SVM | SM |
| | JointRec[85] | CFL, EFA | Sync, CEC | HFL | Bas | N/A | StH | N/A | CNN | SM |
| | FedNewsRec[86] | CFL, EFA | Sync | HFL | DP | N/A | N/A | N/A | DNN, LSTUR | SM |
| DRA | PoCI[95] | N/A | N/A | HFL | HE | N/A | N/A | N/A | LDA | SM |
| SF | FedCoin[94] | CFL, FedAvg | Sync | HFL | Bas | Yes | N/A | TF | NN | SM |
| HM | [84] | CFL, FedAvg | Sync | HFL | DP | N/A | N/A | N/A | DNN | SM |
| | [81] | CFL, FedAvg | Sync | HFL | Enc | N/A | N/A | FATE | DNN | SM |
| | [82] | CFL, FedAvg | Sync | HFL | DP | N/A | N/A | N/A | LR | SM |
| | FedHealth[80] | CFL, FedAvg | Sync | FTL | HE | N/A | MH | N/A | CNN | SM |
| | CBFL[83] | CFL, FedAvg | Sync | HFL | Bas | N/A | StH | N/A | AE | SM |
| LS | FL-RSSF[91] | CFL, FedAvg | Sync | HFL | Bas | N/A | StH | TF | MLP | SM |
| | FedLoc[92] | CFL, FedAvg, ADMM | Sync | HFL | HE | N/A | N/A | N/A | DNN,GP | SM |
| CV | FedVision[96] | CFL, FedAvg | Sync,CEC | HFL | Bas | N/A | N/A | N/A | YOLOv3 | N/A |
| | FL-MEC[97] | CFL, FedAvg | Sync | HFL | Bas | N/A | StH | TF | CNN | SM |
| ST | FL-JPRA[89] | CFL, EFA | Async,Sync CEC | HFL | Bas | N/A | N/A | N/A | GPDPE | SM |
| | FedGRU[90] | CFL, FedAvg | Sync | HFL | Enc | N/A | N/A | PySyft | GRU | SM |
| MPC | F-SVM[93] | CFL, FedAvg | Sync | HFL | Bas | N/A | N/A | N/A | SVM | SM |

Table II presents the research of applying FL into the applications. The abbreviations are as follows: training model (TM), communication efficiency and cost (CEC), federated transfer reinforcement learning (FTRL), differential privacy (DP), gated recurrent unit (GRU), deep deterministic policy gradient (DDPG), TensorFlow (TF), long- and short-term user representation (LSTUR), latent dirichlet allocation (LDA), proof of common interest (PoCI), federated energy demand learning (FEDL), Gaussian process (GP), homomorphic encryption (HE), encryption (Enc), multilayer perceptron (MLP), generalized Pareto distribution parameter estimation (GPDPE), logistic regression (LR), AutoEncoder (AE), synchronous (Sync), asynchronous (Async), self-made (SM), model heterogeneity (MH), statistical heterogeneity (StH), enhanced-FedAvg (EFA), you only look once (YOLO), alternating direction of multipliers method (ADMM), basic (Bas). The 'Bas' in PP column denotes that the application applies the basic privacy preserving built in FL. The 'SM' in Ben column denotes that the application does not use a benchmark, but a self-made benchmark instead.

federated averaging (CDW_FedAvg) algorithm is proposed to solve the data heterogeneity, which considers the distance between positive and negative classes of each client dataset. Ge et al. [73] gave the empirical research results of FL based production line fault prediction. Federated support vector machine (SVM) and federated random forest (RF) algorithms for HFL and VFL are designed respectively. An experimental process is proposed to evaluate the effectiveness of FL and centralized learning algorithms. It is found that there is no significant difference in the performance of between FL and centralized learning algorithms on global test data, random partial test data and estimated unknown Bosch data. Zhang et al. [74] designed an FL method for machinery fault diagnosis based on DL. A dynamic verification scheme based on FL framework is proposed to adjust the model aggregation process adaptively, which ignores the low quality data of some clients.

Furthermore, a self supervised learning scheme is proposed to learn structural information from limited training data. This scheme has dual effects of data augmentation and multi task learning. Experiments on two rotating machinery datasets show that this method provides a promising FL method for fault diagnosis. However, there is still a significant gap between the proposed method and the traditional centralized training method with the Non-IID.

Edge device failures seriously affect the production of industrial products in IIoT. In order to solve this problem, Liu et al. [75] proposed a new communication-efficient on-device FL-based deep anomaly detection framework for sensing time-series data in IIoT. It enables distributed edge devices to train anomaly detection model cooperatively, so as to improve its generalization ability. An attention mechanism-based CNN-LSTM (AMCNN-LSTM) model is proposed to detect anomalies accurately. It uses the CNN module based on attention mechanism to capture important fine-grained features, so as to prevent memory loss and gradient dispersion. It uses LSTM module to accurately and timely detect anomalies. A gradient compression mechanism based on Top-$k$ selection is proposed to improve the communication efficiency and meet the timeliness of industrial anomaly detection.

The digital twin in IIoT maps the running state and behavior of devices to the digital world in real time. By considering the deviation between the digital twin and the actual value of device state in the trust-weighted aggregation strategy, Sun *et al*. [76] quantified the contribution of devices to the global aggregation of FL. The reliability and accuracy of the learning model are improved. Based on deep Q network (DQN), an adaptive calibration method of global aggregation frequency is proposed, which minimizes the loss function of FL under a given resource budget, and realizes the dynamic tradeoff between computing energy and communication energy in time-varying communication environment. In order to further adapt to the heterogeneous IIoT, an asynchronous FL framework was proposed, which eliminates the straggler effect of clustering nodes and improves the learning efficiency through appropriate time-weighted inter-cluster aggregation strategy. This framework determines the clustering frequencies of different clusters through the adaptive frequency calibration based on DQN. Li *et al*. [77] created an FL-based intrusion detection model named DeepFed with CNN and GNU to detect network threats against industrial cyber-physical systems. The designed FL framework allows multiple industrial cyber-physical systems to establish a comprehensive intrusion detection model in a way of privacy protection. A secure communication protocol based on Paillier cryptosystem was designed to keep the security and privacy of model parameters through the training process. The experiments on the data set of a real industrial cyber-physical system show that the model is highly effective in detecting various types of network threats in industrial cyber-physical systems.

Liu *et al*. [33] proposed a learning architecture for cloud robotic system navigation, lifelong federated reinforcement learning (LFRL). LFRL can make the navigation-learning robots use prior knowledge effectively and adapt to the new environment quickly. A knowledge fusion algorithm (KFA) was designed for upgrading the shared model deployed on the cloud, and the transfer methods are introduced. LFRL is consistent with human cognitive science and suitable for cloud robotic system. Liu *et al*. [78] proposed an imitation learning framework for cloud robotic systems with heterogeneous sensor data, called federated imitation learning (FIL). FIL can use the knowledge of other robots in the cloud robotic system to improve the efficiency and accuracy of local robots' imitation learning. In addition, a KFA based on RGB images, depth images and semantic segmentation images was proposed, and a transfer method was introduced in FIL.

In industrial working environment monitoring, it is very important yet difficult to follow the changing trend of the time series monitoring data when they come from different types of sensors and are collected by different companies. FL structure can not only keep the data privacy but also extract and fuse the trend features of time-series monitoring data of multi-sensors. Hu *et al*. [79] considered the conduction model and feature aggregation framework in FL, and proposed a trend following method to put all the fusion features of the multi-sensor time-series monitoring data into the echo state network to realize the multi-sensor electromagnetic radiation intensity time-series monitoring data sampling of the actual mine.

*Healthcare & Medical (HM)*

Protecting highly sensitive information is the shared responsibility of all parties including hospitals, AI companies, and corresponding regulatory agencies. Chen *et al*. [80] proposed the first FTL framework for wearable healthcare - FedHealth. FedHealth can achieve accurate and personalized healthcare without compromising privacy security. Xiong *et al*. [81] established a cross-silo federated drug discovery learning framework based on FATE for predicting drug-related properties and solving the dilemma of small and biased data in drug discovery. Pfohl *et al*. [82] studied the efficacy of centralized learning and FL in private and non-private environments. The clinical prediction tasks are to predict the prolonged length of stay and the mortality rate of thirty-one hospitals. They found that while training in a centralized setting, differential private stochastic gradient descent can be directly applied to achieve a strong privacy boundary, it is much more difficult to do so in a federated setting. Huang *et al*. [83] introduced a community-based federated learning (CBFL) algorithm. The algorithm clusters distributed data into clinically meaningful communities that capture similar diagnoses and geographic locations, and learns a model for each community. Li *et al*. [84] studied the feasibility of applying differential privacy to protect patient data in an FL setting. An FL system was implemented and evaluated for brain tumor segmentation on the BraTS dataset.

*Recommender System (RS)*

Duan *et al*. [85] proposed a joint cloud video recommendation framework based on deep learning - JointRec. It integrates the JointCloud architecture into the mobile IoT to realize joint training among distributed cloud server for video recommendation. Qi *et al*. [86] proposed a FedNewsRec framework to coordinate a large number of users, and jointly train an accurate news recommendation model from the behavior data of these users without uploading raw data. Muhammad *et al*. [87] introduced a federated collaborative filtering (FCF) method for personalized recommendations. This method federates the standard collaborative filtering (CF) with stochastic gradient descent. Hartmann *et al*. [88] introduced an FL system built for use in Firefox. Users can type half a character less to find what they want.

*Smart Transportation (ST)*

Samarakoon *et al*. [89] proposed a distributed, FL-based, joint power and resource allocation (FL-JPRA) framework for enabling ultra-reliable and low-latency vehicular communication. An FL mechanism is proposed in which vehicular users partially estimate the tail distribution with the help of roadside units. Liu *et al.* [90] proposed an FL-based recurrent unit neural network algorithm (FedGRU) for predicting traffic flow.

*Localization Service (LS)*

Because of the low cost and easy implementation of localization based on received signal strength fingerprints (RSSFs), many studies have been conducted. It has promoted the emergence of many commercial applications based on localization services. Ciftler *et al.* [91] proposed a localization technology based on FL and RSSFs (FL-RSSF) to provide privacy-preserving crowdsourcing for localization. A new collaborative positioning and location data processing framework, FedLoc, is proposed, and all the building blocks required to build this framework were reviewed [92]. They put more efforts into the actual user cases of FedLoc and their implementation.

*Mobile Packet Classification (MPC)*

Bakopoulou *et al.* [93] applied a Federated SVM (F-SVM) for Mobile Packet Classification, which allows mobile devices to collaborate and train global models without sharing the original training data. A reduced feature space, HTTP key, is proposed, which limits the sensitive information shared by users.

*Payment in Smart Finance (SF)*

Liu *et al.* [94] proposed a blockchain-based payment system, FedCoin, to enable FL. It can mobilize free computing resources in the community to perform the expensive computing tasks required by the FL incentive plan. FedCoin can correctly determine the contribution of the FL client to the global FL model based on the Shapley value, and has an upper limit on the computing resources required to reach an agreement.

*Data Relevance Analysis (DRA)*

In this information age, the continuous generation of data has brought the problem of finding a needle in a haystack to determine useful data from a bunch of irrelevant data. Doku *et al.* [95] proposed a consensus mechanism called proof of common interest (PoCI) to store the most relevant data found when users interact with mobile devices by combining the trust mechanism of blockchain and FL.

*Object Detection in Computer Vision (CV)*

Through joint learning, the challenge of using image data owned by different organizations to establish an effective visual target detection model is solved.
Liu *et al.* [96] built a FedVision platform, an end-to-end ML engineering platform that supports the easy development of FL-powered computer vision applications. The challenge of using image data owned by different organizations to establish an effective visual target detection model is solved with FL. How to accurately detect and classify targets and perfectly combine the corresponding virtual content with the real world is a major challenge for AR technology. Chen *et al.* [97] proposed a framework combining FL and MEC, FL-MEC, to solve the corresponding challenge.

*Traffic Sign Classification in Smart City (SC)*

The amount of labeled data collected in smart cities is small, and there is a lot of unlabeled data. Albaseer *et al.* [98] proposed a semi-supervised federated edge learning method, called FedSem, to utilize unlabeled data in smart cities. FedSem can use unlabeled data to improve learning performance, even if the ratio of labeled data is low.

*Energy Prediction in Smart Energy (SE)*

Saputra *et al.* [99] proposed a federated energy demand learning method that allows charging stations to share their information without exposing the real dataset. The cluster-based energy demand learning method is applied in charging stations to further improve the accuracy of energy demand prediction.

*Anomaly Detection and Voice Assistant in Smart Building (SB)*

Nguyen *et al.* [100] developed a federated self-learning anomaly detection system for IoT - DÏOT, to use the unlabeled crowdsourcing data captured in the customer's IoT to learn anomaly detection models independently. Leroy *et al.* [101] studied the resource-constrained wake word detector with FL on crowdsourced speech data. Using an adaptive averaging strategy instead of a standard weighted model averaging can greatly reduce the number of communication rounds required to achieve the target performance.

*Collision Detection and Imitation Learning in Autonomous Driving (AD)*

Liang *et al.* [102] presented an online federated reinforcement learning transfer process for real-time knowledge extraction. In this process, all participants will make corresponding actions based on the knowledge of others.

B. Future opportunities in Industry

As illustrated by the FL-transformed manufacturing in Figure 2, FL could be applied to the entire product life cycle. FL also gives small data users (such as SMEs) an opportunity to make full use of intelligence. Specially to our understanding, FL could be seamlessly integrated into the following industrial applications:
- **Product recommendation systems**. In the non-FL setting, manufacturers can only make product recommendations rely on their own sales. Companies should obtain more accurate recommendation services if they utilize FL mechanism to train the recommendation model.
- **Industrial equipment health monitoring**. Modern

industrial equipment is being connected to the Internet via IoT, and their health status can be monitored by big data intelligence. However, few companies have data enough for supporting data intelligence. In this case, industrial companies with similar equipment can apply FL mechanism to harvest federated intelligence for monitoring equipment's health more accurately.

- **AR/VR-guided operations**. AR/VR has been widely used in industries, such as remote operation guidance, virtual assembly and machine operation training. Industrial companies can use FL strategies to train optimal models to improve the accuracy of detecting objects.
- **Precise robotics collaboration**. Traditional RFID-based positioning accuracy is not high. RSSF positioning based on FL can achieve higher accuracy. This FL-enhanced precise positioning can be applied to robotics collaboration.
- **Industrial environmental monitoring**. It is very important yet difficult to track time-series monitoring data on industrial environment collected by different types of sensors and different companies. At the same time, the privacy of data on the operating environment needs to be protected. We can utilize FL strategy to solve such problems.
- **Product defect detection**. DL has a broad application prospect in the field of automatic detection. One of the biggest challenges of applying DL based methods to product defect detection is the lack of data samples for classification task of defect detection. Multiple enterprises that produce similar products can be attracted to join FL to realize sample expansion.
- **Optimal supply chain scheduling**. Traditionally, the data on sales forecast across-regional distributors/industry associations is private. To realize efficient supply chain scheduling, manufacturers can encourage suppliers to participant in FL to extract the optimal model for predicting demand orders, supply quantity, inventory, and supply schedule.
- **Generative product design**. The design data from different companies are only available to themselves for privacy reasons. To shorten the design cycle and reduce design iterations, FL is expected for companies to optimize the generative product design process across enterprises based on the modeling of the human/machine/material resources in each enterprise.
- **Security**. Most of the existing AI intrusion detection schemes for IIoT are designed based on a strong assumption that there are always enough high-quality network attack instances for IIoT [77]. However, in real-world scenarios, a company usually has only a limited number of attack cases, which makes it a great challenge to build a model. In addition, companies are usually reluctant to share such attack instances (including normal behavior instances) with third parties, because these data always involve their highly sensitive information. Intrusion detection schemes based on FL can be used to solve this problem.

## V. CONCLUSION

In this paper, we revisit FL from the perspective of Industry 4.0 emphasizing its application in advancing intelligent manufacturing. To facilitate a common understanding of the FL paradigm, we elaborate and update relevant concepts of the roles, algorithms, tools used in FL, such as learner, organizer, local model, federal model, etc. With the comprehensive survey, the state of the art of FL on fundamental FL research is analyzed from eight topics and further work and challenges are presented. Before reviewing the FL applications in advancing more than thirteen economic sectors, we present the paradigm of FL-transformed manufacturing. Clearly, more attention should be paid on the investigation of integrating FL into Industry 4.0. Meanwhile, we list some industrial areas for IIoT researchers and practitioners into which FL could be seamlessly and immediately integrated. Our other findings are summarized as follows：

- Recently, the attention and research on FL have increased exponentially. However, there is not much research on Industry 4.0 and smart manufacturing. This deserves more attention from the industrial academia and practice on FL.
- The fundamental research corresponding to the recent applications is distributed in the eight areas, and most of them focus on data distribution, model optimization, and privacy protection. However, privacy protection lacks a measurement standard and the suitable quantitative evaluation is missed. We initially present and define the problem in this paper. On the other hand, there are few benchmarks and tool platforms. It can be seen that FL is still in its infancy stage.
- Almost all the surveyed applications are based on CFL. Most of them are based on HFL. Few are based on VFL and FTL, which needs more attentions and efforts in the future.
- The application is increasing in the IIoT, such as fault prediction, device failure detection, cloud robotic system, etc. However, there is huge potential space for FL to accelerate PLCM in the context of IIoT. Other applications mainly fall into categories of healthcare & medical and recommendation systems. Medical care focuses on drug discovery, medical image processing, privacy preservation of electronic health records, and activity recognition. Recommendations include entertainment, news, videos, and automatic text input on the browser.